\documentclass[journal]{IEEEtran}

\usepackage{cite}
\usepackage{amsmath,amssymb,amsfonts}
\usepackage{algorithmic}
\usepackage{graphicx}
\usepackage{textcomp}
\usepackage{xcolor}
\usepackage{bm}
\usepackage{amsthm}
\usepackage[utf8]{inputenc}
\usepackage[T1]{fontenc}
\usepackage{lmodern}
\usepackage[english]{babel}

\usepackage{amsfonts}
\usepackage{tikz}
\usetikzlibrary{positioning,shapes}
\usepackage{multirow}
\usepackage{booktabs}
\usepackage{tabularx}

\usepackage{tabularx}\usepackage[detect-all,per-mode=symbol]{siunitx}
\usepackage{placeins} 
\usepackage{graphicx}
\usepackage{subcaption}

\def\BibTeX{{\rm B\kern-.05em{\sc i\kern-.025em b}\kern-.08em
    T\kern-.1667em\lower.7ex\hbox{E}\kern-.125emX}}

\begin{document}

\title{Reliable and Private Anonymous Routing for Satellite Constellations}

\author{
  Nilesh Vyas \qquad Fabien Geyer \qquad  Svetoslav Duhovnikov \\
  \thanks{N. Vyas, F. Geyer and S. Duhovnikov  are with Airbus Central R\&T, Taufkirchen, 82024 Germany (e-mail: \{nilesh.vyas, fabien.geyer,  svetoslav.duhovnikov\}@airbus.com}
}

\maketitle

\begin{abstract}
\noindent Shared, dynamic network infrastructures, such as dual-use LEO satellite constellations, pose critical threats to metadata privacy, particularly for state actors operating in mixed-trust environments. This work proposes an enhanced anonymity architecture, evolving the Loopix mix-network, to provide robust security and reliability in these volatile topologies. We introduce three primary contributions:  (1) A multi-path transport protocol utilizing $(n, k)$ erasure codes, which is demonstrated to counteract the high link volatility and intermittent connectivity that renders standard mix-networks unreliable. (2) The integration of a computationally efficient Private Information Retrieval (PIR) protocol during route discovery. (3) The introduction of adaptive, centrality-based delay strategies that efficiently mitigate the inherent topological bias of LEO networks, providing a superior anonymity-to-latency trade-off. This mechanism provably prevents metadata leakage at the user-provider directory, mitigating profiling and correlation attacks. We validate this architecture via high-fidelity, packet-level simulations of a LEO constellation. Empirical results show our multi-path transport achieves near-zero message loss, establishing a quantifiable trade-off between reliability and bandwidth overhead. Furthermore, microbenchmarks of the PIR protocol quantify its computational and latency overheads, confirming its feasibility for practical deployment. This work provides a validated blueprint for deployable high-anonymity communication systems, demonstrating the viability of securely multiplexing sensitive operations within large-scale commercial network infrastructures.

\end{abstract}


\maketitle

\section{Introduction}

The architecture of modern communication is increasingly migrating towards a reliance on shared, multi-tenant infrastructures, encompassing terrestrial domains like 6G networks \cite{Eu6G} as well as Non-Terrestrial Networks (NTNs), including Low Earth Orbit (LEO) satellite constellations \cite{EU_Regulation_2023_588, sesar}. In these environments, diverse entities—such as government agencies, commercial enterprises, and private individuals—share the same physical resources \cite{EU_Regulation_2023_588}.

This convergence, while economically and operationally efficient, introduces significant vulnerabilities by inverting the network's foundational design advantages. The direct global accessibility intended for ubiquitous service provides adversaries with a vastly expanded, geographically distributed attack surface. Simultaneously, the stringent architectural imperative for low latency constrains routing options, resulting in deterministic network paths that offer adversaries a high degree of predictability for reconnaissance and attack planning \cite{ICARUSAL}. This vulnerability is particularly acute in dual-use satellite constellations \cite{IRIS, SES} that must securely serve both civilian and military functions, making them high-value targets for state-level adversaries \cite{Cannon2023}.

This elevated threat landscape, compounded by the adoption of standard IP protocols, demands a paradigm shift in security. These networks not only inherit known terrestrial vulnerabilities but also introduce a new class of security challenges unique to their dynamic environment \cite{LeoCyberSec}. Consequently, robust security must extend far beyond content encryption; it is critical to protect communication metadata (i.e., sender/receiver identities and traffic patterns) from sophisticated threats like eavesdropping, jamming, and traffic analysis \cite{Yanev2024, Danezis2015Traffic}. An adversary observing this metadata can infer sensitive information, track activities, and map organizational structures, even if the message content itself is unreadable.

These metadata security concerns are particularly formidable within LEO constellations. Characterized by high node mobility and intermittently available Inter-Satellite Links (ISLs), their topology is in a constant state of deterministic flux. This inherent dynamism is itself a security risk \cite{LEO_SDN}, creating a triad of interconnected challenges that directly threaten network anonymity:

\textit{Path Unreliability:} In a highly dynamic LEO network, even small disruptions can break connectivity, leading to frequent and unpredictable link failures, message loss, and severe performance degradation \cite{yanev2025}. This unreliability is not just a performance issue; it creates a security vulnerability. An adversary can passively observe or actively induce these failures to force users into connection re-establishment, leaking sensitive metadata information during the subsequent connection handshakes \cite{ICARUSAL}.

\textit{Routing Plane Vulnerability:} The highly dynamic topology necessitates a sophisticated, and often centralized, routing and control plane to manage the constant path changes. This management infrastructure, which calculates and distributes routes for the entire network, becomes a critical vulnerability. An adversary capable of observing or compromising this control plane (e.g., a ground station controller) gains access to network-wide flow information. This enables powerful attacks that can correlate traffic patterns to disclose the identity and location of senders or receivers, bypassing data-plane encryption \cite{surveyroutingAC}.

\textit{Topological Bias and Centralization:} The periodicity and regularity of LEO orbits, while predictable, create non-uniform traffic patterns \cite{Leoddos}. This topological bias creates choke points, such as key nodes over polar regions that concentrate traffic. These nodes become high-value targets for adversaries conducting large-scale surveillance or launching Distributed Denial-of-Service (DDoS) attacks \cite{Leoddos}.

This triad of challenges renders existing anonymity networks solutions \cite{Piotrowska20} ineffective for this new environment. Connection-oriented, source-routed designs like onion routing \cite{Reed1998} are fundamentally ill-suited; their directory-based circuits cannot withstand the high topological volatility, and their reliance on a discoverable network view creates a critical metadata vulnerability. While DC-Nets \cite{Chaum1988} and message-based mix-networks \cite{Chaum1981} like Loopix \cite{Piotrowska2017}, I2P's Garlic Routing \cite{Freedman2000}, and Dandelion \cite{bojja2017dandelion}, Crowds \cite{Reiter1998} offer a more promising paradigm that is resilient to link failures, they introduce significant latency overheads. More importantly, they do not solve the underlying vulnerabilities of a compromised routing plane or the inevitable traffic centralization at topological choke points.

To address these challenges, this paper proposes and evaluates a holistic architecture that extends the Loopix mix-network to provide practical, reliable, and metadata-private communication in dynamic LEO networks. By systematically solving for unreliability, routing leaks, and topological bias, our design enables the trusted, anonymous integration of multiple entities on a shared, time-variant infrastructure. Our primary contributions are:
\begin{itemize}
    \item \textit{Solving Path Unreliability:} We design and integrate a resilient communication mechanism using $(n,k)$ erasure codes over multiple, diverse network paths. This provides robustness against message loss from topological instability, preventing the metadata leaks caused by forced connection re-establishments.
    \item \textit{Mitigating Routing Plane Vulnerability:} We apply a privacy-preserving route discovery protocol based on computational Private Information Retrieval (PIR). This allows clients to fetch dynamic routing data from the network's control plane without revealing their queries or communication interests.
    \item \textit{Counteracting Topological Bias:} We design and evaluate adaptive delay strategies that actively neutralize traffic centralization. We demonstrate that basing mix delays on network centrality, rather than a baseline uniform strategy, effectively diffuses traffic patterns at predictable choke points and achieves stronger anonymity for a given latency.
    \item \textit{Performance and Feasibility Validation:} We present a detailed performance evaluation via high-fidelity simulation of a 631-satellite OneWeb-like constellation. This validation quantifies the practical trade-offs between anonymity, reliability, and overhead (computation, bandwidth, latency).
\end{itemize}

We demonstrate that our architecture is practically feasible, achieving near-perfect reliability and private route queries against a realistic database. This solution provides a viable path toward building the trust and integrity required for secure, anonymous routing in shared, time-variant LEO networks.

\section{Related Works}

\subsection{Loopix: A Modern Low-Latency Mixnet}
Loopix was introduced as a next-generation anonymous communication system designed to resolve this tension, offering the strong anonymity properties of a mixnet with significantly lower latency \cite{Piotrowska2017}. It is a message-based system built on a stratified (layered) network of mixes and semi-trusted providers that manage user access and store messages for offline recipients.

Loopix achieves its security goals through two primary mechanisms: (1) \textit{Poisson Mixing:} Instead of using synchronized batches, each mix node independently delays incoming messages according to a random value drawn from an exponential distribution. This continuously de-links the timing of input and output messages, providing robust protection against a global passive adversary capable of observing all network traffic. (2) \textit{Continuous Cover Traffic:} The system injects a constant stream of cover traffic to obfuscate real user patterns. This includes drop messages, which are discarded at their destination, and loop messages, which are routed through the network and back to the original sender.

These loop messages are a key innovation. They serve a dual purpose: they provide cover for real messages and, more importantly, they function as an active defense mechanism. Senders can detect if their loops fail to return, signaling a potential message-dropping or $n-1$ (isolation) attack by a malicious node on the path.

Loopix's design successfully provides quantifiable anonymity with practical end-to-end delays on the order of seconds, offering a scalable and robust alternative to traditional high-latency systems.

\subsection{Research Gap: Loopix in Dynamic Networks}

While the message-based paradigm of Loopix \cite{Piotrowska2017} is a promising foundation, its core architecture is predicated on a crucial assumption: a quasi-static network topology. This assumption makes it directly vulnerable to the fundamental LEO network challenges previously identified.

First, Loopix's source-routing mechanism is incompatible with the path unreliability challenge. In a highly dynamic LEO environment, pre-defined routes will frequently break, leading to the exact message loss and performance degradation. This failure mode, in turn, triggers the associated metadata leak by forcing users to repeatedly re-establish connections, which aids deanonymization efforts \cite{ICARUSAL}. Although prior work on mixnet reliability has explored solutions like mix groups \cite{Zhuang2005} or threshold cryptography \cite{Yusheng2020}, these often impose significant latency and computational overhead. We build on the more recent concept of network coding \cite{Zhihan2022}, proposing a lightweight erasure-coding scheme (Section \ref{sec:erasure_codes}) to achieve route reliability without this complexity.

Second, this high rate of path failures directly exacerbates the routing plane vulnerability. The constant need to discover new, valid paths forces clients to frequently query the network's directory service. The baseline Loopix design does not sufficiently protect this lookup process, creating a high-frequency metadata leak. An adversary monitoring this directory can observe who is querying, when they query, and how often. This query pattern becomes a potent side-channel, allowing an adversary to correlate activity with specific users or events. This vulnerability confirms the critical need for a private route discovery mechanism, which we address using Private Information Retrieval (PIR) in Section \ref{sec:anonymous_route_discovery}.

\section{A Resilient and Privacy-Preserving Architecture}
This section details the design of the proposed system, which enhances a baseline Loopix network with mechanisms for reliable transmission and private route discovery.

\subsection{System Overview and Threat Model}
The system consists of clients, providers, a network of Loopix mix nodes, and a Route Information Directory (RID). Clients connect to providers to send and receive messages through the mix-network. To construct source-routed paths, clients must query the RID, which maintains a view of the current network topology.

The threat model assumes a \textit{global passive adversary} who can observe all traffic flowing between all entities in the network. This adversary can perform traffic analysis, timing analysis, and other correlation attacks but is assumed to be computationally bounded and cannot break the underlying cryptographic primitives (e.g., AES, public-key encryption). The model also considers malicious node operators, where the adversary may control a fraction of the mix nodes and/or the RID server itself. 

In addition to these adversarial threats, our model explicitly accounts for the inherent environmental threats of the LEO network: high topological volatility, intermittent link connectivity, and the resulting route staleness. While not adversarial in nature, this environmental volatility creates the path unreliability and high message loss that our contribution is designed to mitigate. An adversary can also passively exploit this volatility, as connection failures may force users to re-establish connections and leak deanonymizing metadata. While we also consider malicious node operators who perform traffic analysis, this work focuses on defending against a global passive adversary. Active attacks, such as targeted jamming or Denial-of-Service (DoS) on network choke points, are considered out of scope for our anonymity analysis

The primary security goal is to maintain sender-receiver unlinkability, meaning the adversary should not be able to determine which pairs of users are communicating, even in the face of dynamic network changes and malicious colluding entities.

\subsection{Reliable Multi-Path Transmission with Erasure Coding}
\label{sec:erasure_codes}
To address the challenge of path unreliability in dynamic networks, the system integrates a Forward Error Correction (FEC) mechanism based on erasure codes. This allows a message to be successfully delivered even if some of its constituent packets are lost in transit.

\textit{Protocol Flow:} The process involves the following steps:
\begin{enumerate}
    \item \textit{Fragmentation:} When a client wishes to send a message, its provider first splits the message into $k$ data chunks of a fixed size.
    \item \textit{Encoding:} The provider then applies a systematic $(n, k)$ Reed-Solomon erasure code to these $k$ data chunks to generate $p = n - k$ parity chunks. This results in a total of $n$ chunks.
    \item \textit{Path Selection and Encapsulation:} The provider queries the RID (using the private mechanism described below) to obtain topology information. It then selects $n$ diverse, ideally node-disjoint, paths through the Loopix mix-network. Each of the $n$ chunks is encapsulated in a separate, independent Sphinx packet, each routed along one of the selected paths.
    \item \textit{Transmission:} The provider injects all $n$ Sphinx packets into the mix-network.
    \item \textit{Reconstruction:} The recipient's provider collects the incoming chunks. Due to the properties of Reed-Solomon codes, it can reconstruct the original message as soon as it has received any $k$ of the $n$ transmitted chunks. This design can tolerate the complete failure of up to $p$ paths without any loss of data.
\end{enumerate}

This mechanism provides two distinct benefits. The primary benefit is increased reliability, as it makes communication robust against the link failures and node outages common in dynamic environments. A secondary benefit is enhanced anonymity, as splitting a single logical message across multiple, uncorrelated paths adds noise to the network, making it even more difficult for an adversary to perform traffic analysis and correlate communication flows.

\begin{figure*}[htbp]
	\centering
	\begin{tikzpicture}[
		line cap=round,
		mixnode/.style={rectangle, fill=black!20, draw, very thick, minimum size=5mm, rounded corners=2pt},
		providernode/.style={rectangle, fill=black!50, draw, very thick, minimum size=5mm, rounded corners=2pt},
		usernode/.style={rectangle, draw, very thick, minimum size=5mm, rounded corners=2pt},
		emixnodes/.style={dashed, gray},
		bpath/.style={blue,line width=3pt,-latex},
		rpath/.style={red,line width=3pt,-latex},
		ypath/.style={green,line width=3pt,-latex},
		]
		
		\node[mixnode] (mn00) {};
		\node[mixnode] (mn10) [below=of mn00] {};
		\node[mixnode] (mn20) [below=of mn10] {};
		
		\node[mixnode] (mn01) [right=of mn00] {};
		\node[mixnode] (mn11) [below=of mn01] {};
		\node[mixnode] (mn21) [below=of mn11] {};
		
		\node[mixnode] (mn02) [right=of mn01] {};
		\node[mixnode] (mn12) [below=of mn02] {};
		\node[mixnode] (mn22) [below=of mn12] {};
		
		\node [cloud, draw, gray, cloud puffs=11, cloud puff arc=100, aspect=1.2, inner ysep=4em] (mncloud) at (mn11.center)
		{};
		\node (mnlabel) [below=5pt of mncloud] {Mix Nodes};
		
		\node[providernode] (pn0) [left=50pt of mn10] {};
		\node[align=center] (pn0label) [below=5pt of pn0] {Provider\\ Node};
		\node[providernode] (pn1) [right=50pt of mn12] {};
		\node[align=center] (pn1label) [below=5pt of pn1] {Provider\\ Node};
		
		\node[usernode] (u0) [left=of pn0] {};
		\node[align=center] (u0label) [below=5pt of u0] {User\\ Node};
		\node[usernode] (u1) [right=of pn1] {};
		\node[align=center] (u1label) [below=5pt of u1] {User\\ Node};
		
		\draw[emixnodes] (pn0) -- (mn00) -- (mn01) -- (mn02) -- (pn1);
		\draw[emixnodes] (pn0) -- (mn10) -- (mn11) -- (mn12) -- (pn1);
		\draw[emixnodes] (pn0) -- (mn20) -- (mn21) -- (mn22) -- (pn1);
		\draw[emixnodes] (mn00) -- (mn11) -- (mn02);
		\draw[emixnodes] (mn00) -- (mn21) -- (mn02);
		\draw[emixnodes] (mn10) -- (mn01) -- (mn12);
		\draw[emixnodes] (mn10) -- (mn21) -- (mn12);
		\draw[emixnodes] (mn20) -- (mn01) -- (mn22);
		\draw[emixnodes] (mn20) -- (mn11) -- (mn22);
		
		\draw[line width=3pt,-latex] (u0) to (pn0);
		
		\draw[ypath] (pn0) to (mn10);
		\draw[ypath] (mn10) to (mn21);
		\draw[ypath] (mn21) to (mn02);
		\draw[ypath] (mn02) to (pn1);
		
		\draw[bpath] (pn0) to (mn00);
		\draw[bpath] (mn00) to (mn01);
		\draw[bpath] (mn01) to (mn12);
		\draw[bpath] (mn12) to (pn1);
		
		\draw[rpath] (pn0) to (mn20);
		\draw[rpath] (mn20) to (mn11);
		\draw[rpath] (mn11) to (mn22);
		\draw[rpath] (mn22) to (pn1);
		
		\draw[line width=3pt,-latex] (pn1) to (u1);
	\end{tikzpicture}
	\caption[Illustration of FEC in Loopix]{Illustration of FEC in Loopix, where three different disjoint paths are taken through the network. The provider nodes are in charge of splitting and reconstructing the data from the user nodes using network coding.}
	\label{fig:loopix_fec}
\end{figure*}
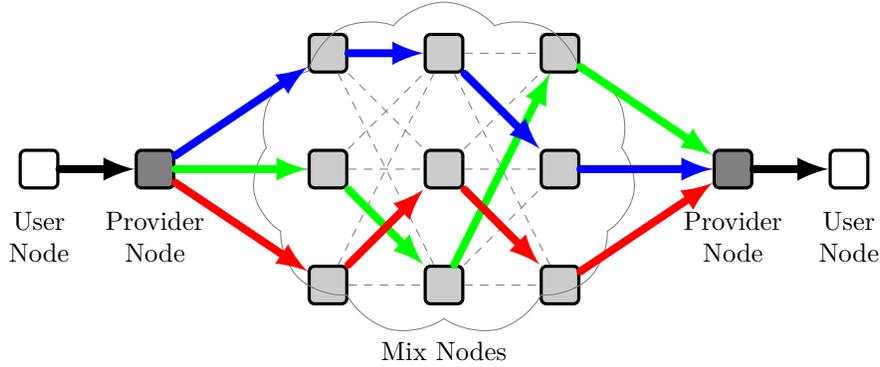

This concept is illustrated in Figure \ref{fig:loopix_fec} and its performance will be evaluated in Section \ref{sec:evalfecperf}.

\subsection{Anonymous Route Discovery with Homomorphic Encryption}
\label{sec:anonymous_route_discovery}

To close the metadata leakage channel during route discovery, the system employs a single-server PIR protocol based on on homomorphic encryption. This allows a client to retrieve a specific route from the RID without the RID learning which route was requested. It is critical to note that this design differs from directory-based systems like Tor \cite{reed1998onion, Piotrowska2017}. Instead of querying for a list of mix-nodes (as we assume all satellites can be mixes), the client queries for a specific end-to-end physical path from the LEO topology database. This query for a specific path is the primary metadata leak we aim to prevent. 

\textit{Cryptographic Primitive: The BFV Scheme:} The protocol is built upon the BFV somewhat homomorphic encryption scheme \cite{bfv}. BFV is based on the Ring Learning With Errors (RLWE) problem and supports a limited number of homomorphic operations on encrypted data. Crucially for PIR, it allows for the addition of two ciphertexts and the multiplication of a ciphertext by a plaintext. A ciphertext encrypting a message $m_1$ can be added to a ciphertext encrypting $m_2$ to produce a ciphertext encrypting $m_1 + m_2$. Similarly, a ciphertext encrypting $m_1$ can be multiplied by a plaintext $m_2$ to yield a ciphertext encrypting $m_1 \times m_2$.

\textit{PIR Protocol Flow:} The interaction between the client and the RID follows these steps:
\begin{enumerate}
    \item \textit{Setup:} The RID server organizes its database of network routes (e.g., a list of all possible paths of a certain length) into a one-dimensional array, $DB$, of size $N$. Each element $DB_i$ is a plaintext representation of a route.
    \item \textit{Query:} The client, wishing to retrieve the route at index $idx$, generates a public/private key pair $(pk, sk)$. It then constructs a query vector $q$ of length $N$. For each position $i$ from 0 to $N-1$, the client computes $q_i = Enc(pk, 1)$ if $i = idx$, and $q_i = Enc(pk, 0)$ otherwise. This query vector $q$, which consists of $N$ ciphertexts, is sent to the RID server.
    \item \textit{Answer:} The server receives the query vector $q$. Without knowing $idx$, it computes a single resulting ciphertext, $a$, by taking the homomorphic dot product of the database and the query: $a = \sum_{i=0}^{N-1} (DB_i \cdot q_i)$. Due to the homomorphic properties of BFV and the structure of the query vector, this computation results in $a = Enc(pk, DB_{idx})$. The server sends the single ciphertext $a$ back to the client.
    \item \textit{Extract:} The client receives $a$ and decrypts it using its secret key $sk$ to recover the desired route, $DB_{idx}$.
\end{enumerate}

\textit{Optimization with Ciphertext Packing:} The naive protocol described above is impractical as it requires the client to send a query whose size is proportional to the entire database. To make this feasible, the implementation uses a technique called ciphertext packing, which is native to many RLWE-based schemes like BFV. Using the Chinese Remainder Theorem (CRT), it is possible to pack thousands of plaintext values into a single plaintext polynomial, which can then be encrypted into a single ciphertext. Homomorphic operations on these packed ciphertexts act in a Single Instruction, Multiple Data (SIMD) fashion, applying the operation element-wise across all packed values. This allows the client to pack its query vector of 0s and 1s into a much smaller number of ciphertexts, drastically reducing the communication cost of the query phase and making the protocol practical.

\subsection{Integration with the Network Layer}

The abstract concepts of source-routed paths and service functions can be efficiently implemented using a programmable data plane. \textit{Segment Routing with MPLS (SR-MPLS)} is a suitable candidate technology for this purpose \cite{bashandy2019srmpls}. In an SR-MPLS framework, each satellite or ground node in the network can be assigned a unique \textit{Node SID}. The specific Loopix mix function running on a node can be advertised as a \textit{Service SID} \cite{clad2024service}. A complete, source-routed path through the mix-network can then be encoded by the provider as a stack of MPLS  labels (representing the SID) and prepended to the packet. Intermediate routers in the underlying network simply perform efficient label-swapping operations, while only the designated mix nodes perform the more computationally intensive cryptographic processing. This approach is well-suited for the resource-constrained nature of satellite nodes.

\section{Theoretical Models of Anonymity}
The strength of an anonymity system is quantified using Shannon entropy \cite{shannon2001mathematical}, which measures an adversary's uncertainty about the communicating parties. Our analysis is grounded in two complementary information-theoretic models. The first model focuses on the micro-level dynamics of anonymity within a single mix node, while the second provides a macro-level view of end-to-end path anonymity across the network.

\subsection{Incremental Entropy at a Mix Node}
For a discrete random variable $X$ representing the possible senders of a message, the entropy $H(X)$ quantifies the adversary's uncertainty about the true sender. It is defined as $$H(X) = -\sum_{x \in X} p(x) \log_2 p(x),$$ where $p(x)$ is the probability that sender $x$ is the originator. A higher entropy value corresponds to a stronger anonymity guarantee.

The Poisson mix strategy can be modeled as a pool mix, where incoming messages form a pool of indistinguishable messages. The anonymity provided by such a mix can be computed incrementally. Following the model in \cite{Piotrowska2017}, the entropy $H_t$ after a message is sent from a mix at time $t$ can be calculated based on the state of the mix at time $t-1$:
\begin{equation}
H_{t}=H\left(\left\{\frac{k}{k+l},\frac{l}{k+l}\right\}\right) + \frac{k}{k+l}\log_2 k+\frac{l}{k+l}H_{l-1},
\label{eq:entropy}
\end{equation}
where $l$ is the number of messages in the mix pool from the previous round, $H_{l-1}$ is the entropy of that pool, and $k$ is the number of new messages that have arrived since the last departure. This model is particularly useful for simulating the evolution of anonymity over time and understanding the impact of system parameters like traffic load and delay strategy.

This formula captures the two sources of uncertainty for an adversary. The term $H(\{\frac{k}{k+l},\frac{l}{k+l}\})$ represents the adversary's uncertainty about whether the outgoing message is one of the $k$ new arrivals or one of the $l$ old messages. The remaining terms are the weighted average of the uncertainty within each of those sets. This incremental model allows for the efficient simulation of entropy propagation through the entire network. When a message traverses a corrupt node, no new entropy is generated, and the probability distribution associated with the message is passed through unchanged. This model provides the theoretical basis for the simulation results presented in this section, particularly for analyzing the system's resilience to node compromise.

\subsection{Path Anonymity and Resilience to Node Compromise}
To analyze the system's resilience to a static adversary who has compromised a fraction of the network, we adopt a model of \textit{path anonymity} based on work in intermittently connected networks \cite{sakai2015analysis, kate2007anonymity}. This model quantifies the adversary's uncertainty about the true end-to-end path taken by a message. The anonymity set, $\phi$, is the set of all possible acyclic paths of a given length $\eta$.

The maximum entropy, $H_{max}$, occurs when no nodes are compromised. Assuming a total of $n$ nodes, the number of possible acyclic paths of length $\eta$ is given by the permutation $P(n, \eta) = \frac{n!}{(n-\eta)!}$. If each path is equally likely, the maximum entropy is:
\begin{equation}
H_{max} = -\log_2\left(\frac{(n-\eta)!}{n!}\right)
\end{equation}

When an adversary compromises $c$ nodes in the network, the anonymity set shrinks. The expected number of compromised nodes, $c_o$, on a random path of length $\eta$ can be modeled using a binomial distribution:
\begin{equation}
c_o = E[Y] = \sum_{i=1}^{\eta} i \binom{\eta}{i} \left(\frac{c}{n}\right)^i \left(1-\frac{c}{n}\right)^{\eta-i}
\end{equation}
where $Y$ is the random variable for the number of compromised nodes on the path. For each of the $c_o$ compromised nodes, the adversary learns the next hop, reducing the uncertainty. The path anonymity $D(\phi')$, defined as the ratio of the remaining entropy to the maximum entropy, can be approximated as:
\begin{equation}
D(\phi') = \frac{H(\phi')}{H_{max}} \approx 1 - \frac{c_o}{\eta}
\label{eq:path_anonymity}
\end{equation}
This simplified model predicts that path anonymity degrades linearly with the expected fraction of compromised nodes on the path. %It provides the theoretical foundation for analyzing the resilience results presented in Figure \ref{fig:node_compromise}.

\section{Evaluation Methodology}
\label{sec:evaluation}

This section presents the simulation setup, outlines the fundamental challenges to achieving robust anonymity in a dynamic LEO satellite network, and evaluates the performance of our proposed mechanisms, demonstrating how they address these challenges and quantifying the relevant performance trade-offs.

\subsection{Simulation Setup}

\paragraph{Scenario}
We use OneWeb as an exemplary satellite constellation throughout this evaluation.
More specifically, we use the OneWeb constellation as deployed in space using the ephemerids published by CelesTrak from April 30, 2024.
This accounts for a total of 631 satellites.
The constellation is visualized later in Figure \ref{fig:sat-globe-centrality}.

We simulated a total of 50 end-users (Loopix provider nodes) randomly placed on Earth between the latitudes \SI{-55}{\degree} and \SI{55}{\degree}.
We restricted the end-users to these latitudes in order to take into account where the majority of the world population lives.

\paragraph{Loopix network simulations}
For our network simulations of Loopix showed in this section, we ran multiple simulations (between \num{10000} and \num{25000} runs), each run for a total of \SI{300}{\second}.

\paragraph{Hardware}
All benchmarks were executed on a server equipped with an AMD EPYC 7702P 64-Core Processor, providing a consistent high-performance environment for measuring computational costs.

\paragraph{Implementation}
The core anonymity system was implemented as a custom simulator in the Go programming language.
It utilized standard cryptographic libraries, including \texttt{crypto/ecdh} with the X25519 curve for key exchange and \texttt{golang.org/x/crypto} for XSalsa20 and Poly1305 for symmetric encryption and authentication. The erasure coding component was built using an open-source Reed-Solomon library. The PIR protocol was implemented  using Microsoft SealPIR / SEAL (C++) \cite{seal}: the core cryptographic engine, SEAL, provides the fundamental HE operations (encryption, decryption, homomorphic addition/multiplication), while SealPIR offers higher-level abstractions for PIR schemes.

%Despite using a powerful CPU for our simulations, simulating 631 mix nodes and 50 provider nodes in parallel and in real time proved to be difficult once reaching a certain traffic rate. Long delays were observed, mainly coming from the CPU bottleneck. To mitigate this, we used a virtual clock, implemented using an event loop following a discrete time event simulation, a traditional technique also found in other simulators (eg. ns3 or OMNeT++).

\paragraph{PIR Parameters}
After extensive tuning, the following stable and high-performance parameters were chosen:
\begin{itemize}
    \item \textit{Scheme}: \texttt{BFV}, an HE scheme well-suited for integer arithmetic.
    \item \textit{\texttt{poly\_modulus\_degree} (N)}: \texttt{4096}. This parameter determines the size of the ciphertexts and the number of slots available for data packing. \texttt{4096} provided the best balance, offering fast server computation.
    \item \textit{\texttt{plain\_modulus\_bits} (logt)}: \texttt{16}. This defines the size of the space for the data. A 16-bit space is sufficient for the satellite IDs and allows the library to use smaller, faster prime numbers for its calculations.
\end{itemize}

\paragraph{Dataset}
The system used OneWeb constellation as an unweighted graph of 631 nodes (satellites) and their communication links. The pre-processing script computes the shortest path with at least 3 hops between every unique pair of satellites. The resulting collection of \num{191577} paths forms a complete routing table for the network, which serves as the server's database for the PIR simulation.

\subsection{Challenges of Anonymity in LEO Satellite Networks}
Achieving reliable and unlinkable communication in a LEO satellite constellation presents unique challenges not found in terrestrial networks. The constant, high-velocity movement of nodes creates a highly volatile environment that directly impacts routing and anonymity.

\subsubsection{Topological Volatility and Route Staleness}
\begin{figure}[ht!]
    \centering
    \includegraphics[width=\columnwidth]{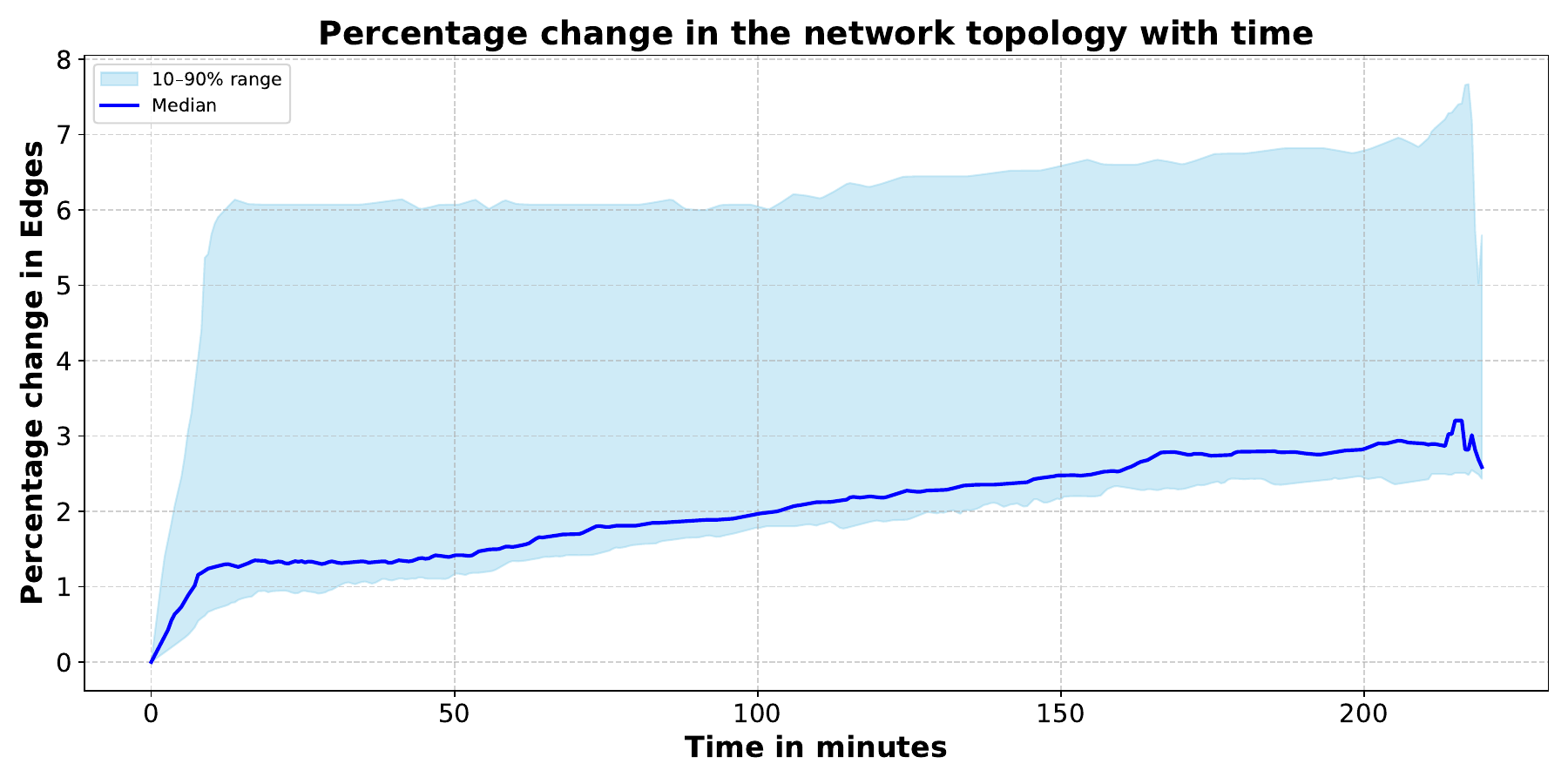}
    \caption{Percentage change in network topology edges over time. The solid line represents the median, and the shaded area denotes the 10–90\% distribution range.}
    \label{fig:topo_change_mins}
\end{figure}

The constant movement of LEO satellites results in a continuously changing network topology. Inter-satellite links are constantly forming and breaking as satellites move in and out of range. Over a longer period, as shown in Figure \ref{fig:topo_change_mins}, the network exhibits a persistent change (median change upto 3\% in 200 mins), confirming that the topology is in a constant state of flux.

\begin{figure}[ht!]
    \centering
    \includegraphics[width=\columnwidth]{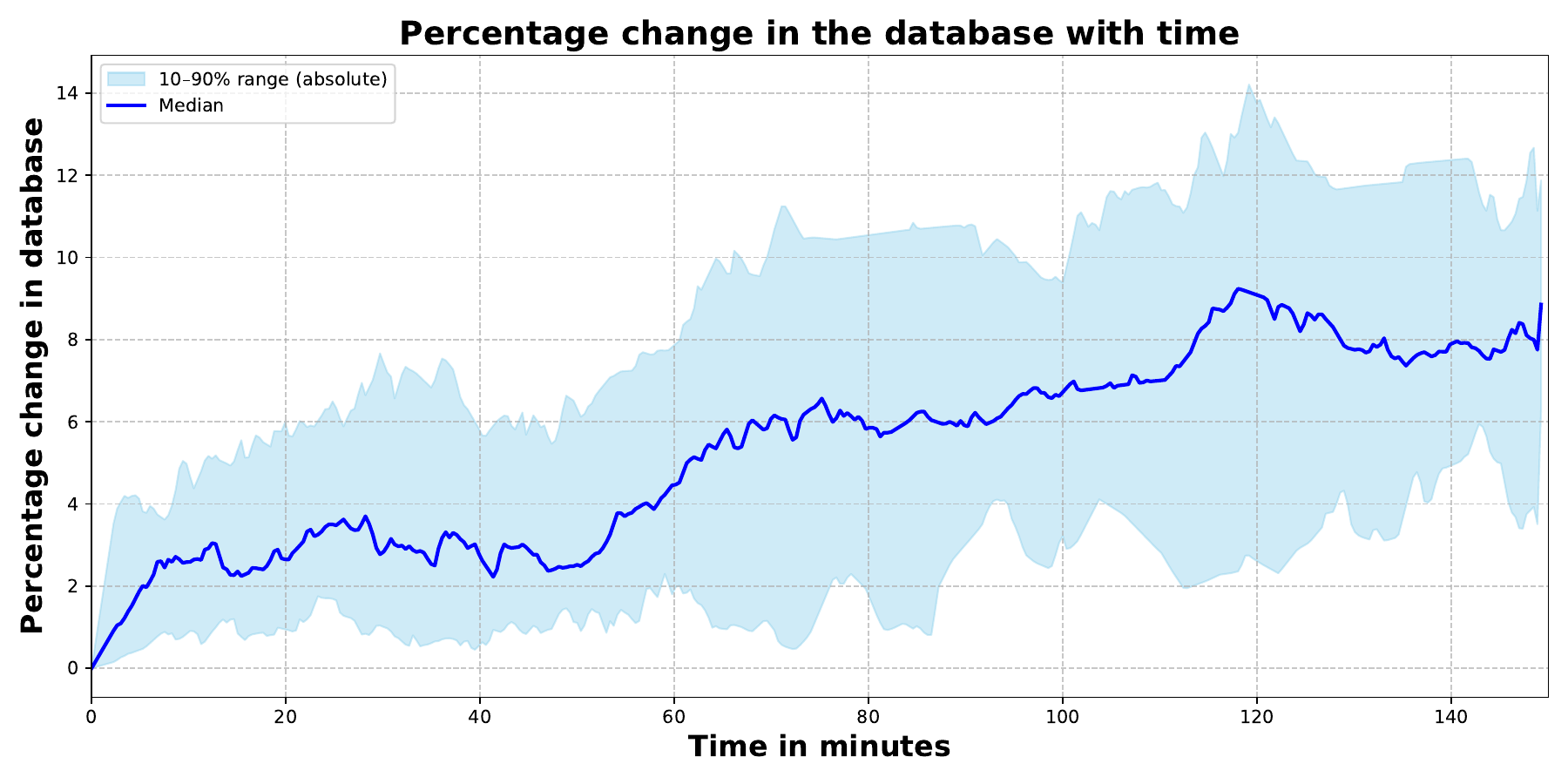}
    \caption{Median percentage change in the database over time. The blue line indicates the median and the light blue region denotes the 10–90th percentile interval}
    \label{fig:db_percent_change}
\end{figure}

This instability has two major consequences:
\begin{enumerate}
    \item \textit{Path Unreliability:} A path that is valid at the time of selection may cease to exist before a message has fully traversed it, leading to high rates of packet loss.
    \item \textit{Route Database Staleness:} The database of valid routes that clients must query is also highly dynamic. As shown in Figure \ref{fig:db_percent_change}, the route in the database changes continuously over time (upto 8\% median change in 120 mins). This means a client's cached routing information becomes stale very quickly, necessitating frequent and private updates to maintain connectivity and avoid metadata leaks. 
\end{enumerate}

\subsubsection{Non-Uniform Traffic Distribution}
\begin{figure}[ht!]
	\centering
	\includegraphics[width=.8\columnwidth]{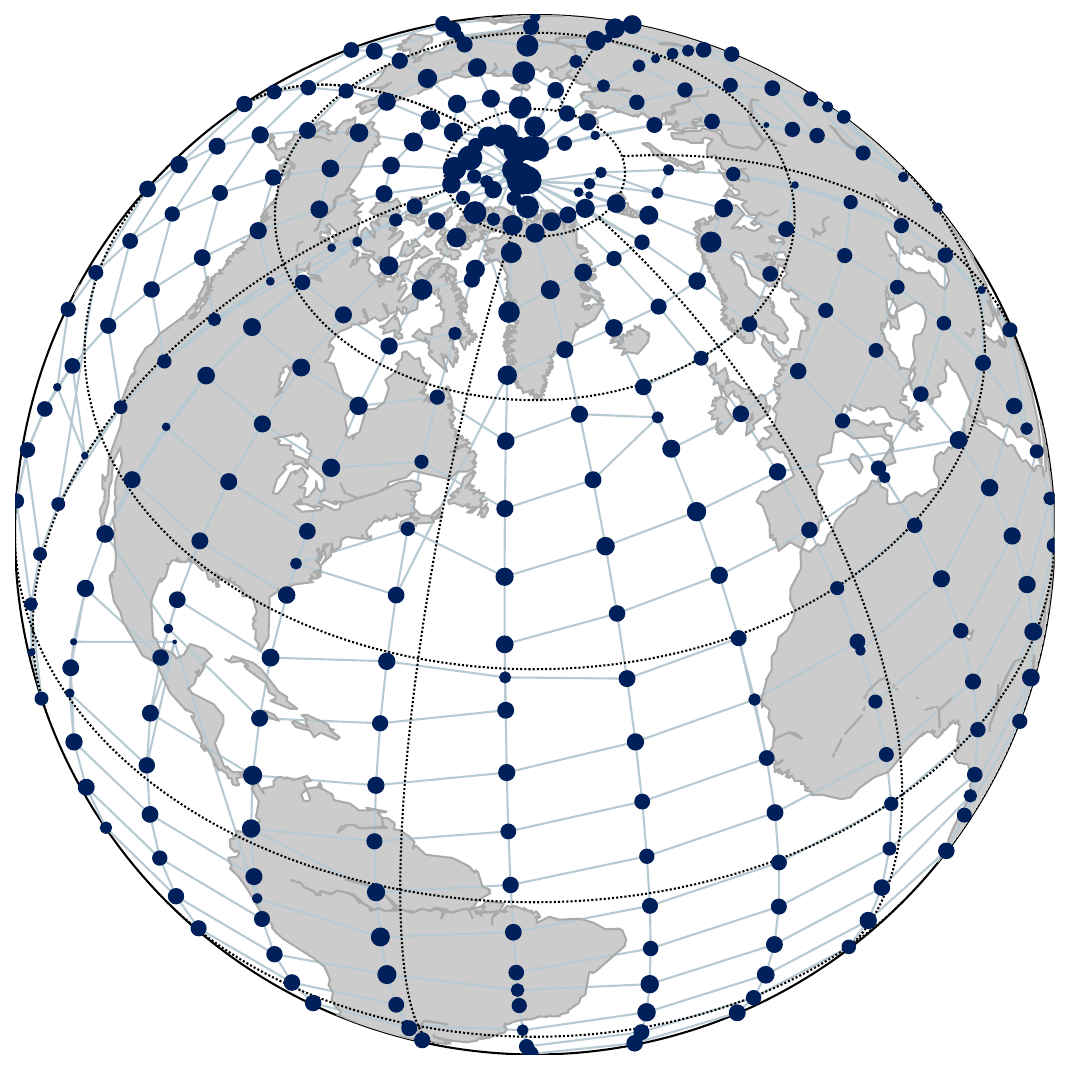}
	\caption{Visualization of node betweeness centrality. Each blue dot represents a satellite, with its size representing its betweeness centrality.}
	\label{fig:sat-globe-centrality}
\end{figure}

\begin{figure}[ht!]
	\centering
	\includegraphics[width=\columnwidth]{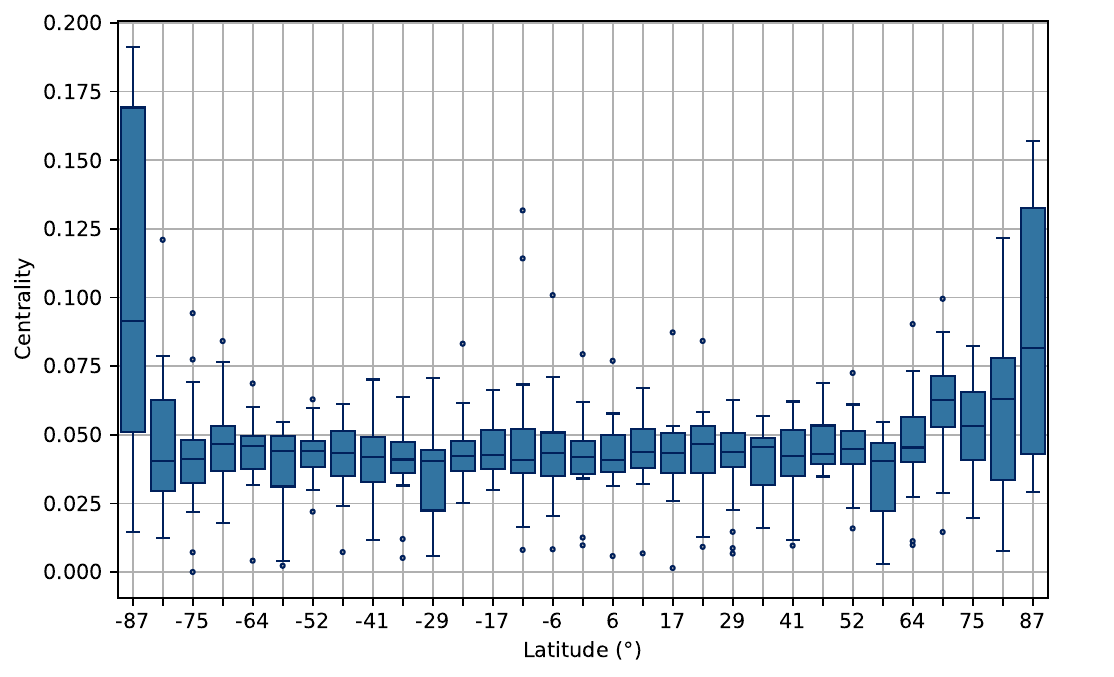}
	\caption{Visualization of node betweeness centrality. Each blue dot represents a satellite, with its size representing its betweeness centrality.}
	\label{fig:sat-latitude-centrality}
\end{figure}

An effective anonymity network should distribute traffic as uniformly as possible to prevent adversaries from identifying high-traffic choke points. However, the inherent topology of a satellite constellation leads to a non-uniform distribution. As shown in Figure \ref{fig:sat-globe-centrality} and \ref{fig:sat-latitude-centrality}, nodes located near the poles have a significantly higher probability of being included in a random path due to the nature of polar orbits. This topological bias means these polar nodes can become performance bottlenecks and are more valuable targets for an adversary seeking to compromise the network.

\section{Performance Analysis: Private Route Discovery in a Dynamic Environment}
\label{Performance PIR}
The practicality of the PIR protocol for satellite routing hinges on optimizing computational latency and architectural design. We implement a high-performance PIR scheme using the Microsoft SealPIR library, which provides high-level abstractions for the Brakerski-Fan-Vercauteren (BFV) homomorphic encryption scheme.

\subsection{Architectural Strategies} 
To address the latency challenges of retrieving paths from a global constellation database, we evaluated four distinct architectural configurations. Preliminary simulations established a maximum path length of 27 hops between any two satellites. To facilitate efficient retrieval, the system employs a public lookup table as a translation layer, mapping unique source-destination pairs to specific row indices. The retrieval pipeline proceeds as follows: the client identifies the target row index via the lookup table, submits it as a query, and decrypts the server's response to reconstruct the path.

\subsubsection{Scenario 1: Single-Server Baseline (Unbatched)}
To establish a worst-case performance baseline, we benchmarked single-server, unbatched configurations across increasing database scales. While effective for smaller paths, attempts to process the full constellation database (27 hops, 191,577 paths) proved computationally infeasible. The unoptimized homomorphic operations caused the server (equipped with an AMD EPYC 7702P 64-Core Processor) to exhaust available memory and terminate abnormally. Consequently, we restricted the baseline scalability analysis to a maximum of 12 hops to ensure stable execution.

\begin{table*}[htbp]
\centering
\caption{Detailed performance breakdown of PIR architectures (Build, Reply, Decrypt, and Total times). \textit{Scenario 1:} Single-Server Baseline (5 hops). \textit{Scenario 2a:} Single-Server Batched  \textit{Scenario 2b:} Single-Server optimized packing.  \textit{Scenario 3:} Parallel servers. All times in milliseconds.}
\label{tab:pir_detailed}
\resizebox{\textwidth}{!}{% Resize to fit width if necessary
\begin{tabular}{@{}llrrrrrrr@{}}
\toprule
\textbf{Configuration} & \textbf{Stage} & \textbf{2 Cores} & \textbf{4 Cores} & \textbf{8 Cores} & \textbf{16 Cores} & \textbf{32 Cores} & \textbf{48 Cores} & \textbf{64 Cores} \\ \midrule

\multirow{4}{*}{\textbf{Scenario 1}} & Build Time & 71.76 & 62.07 & 45.90 & 41.09 & 70.38 &  38.581 &  39.27 \\
 & Reply Time & 110745.00 & 55291.60 & 36870.20 & 37013.60 & 37003.60&  37562.6  & 36971.30  \\
 & Decrypt Time & 34.67 & 20.31 & 12.69 & 19.85 & 14.93 &  14.6348 & 14.14 \\ 
 & \textbf{Total Time} & \textbf{110852.00} & \textbf{55373.90} & \textbf{36928.80} & \textbf{37074.50} & \textbf{37088.90} &  \textbf{37615.8} &  \textbf{37024.70} \\ \midrule

\multirow{4}{*}{\textbf{Scenario 2a}} & Build Time & 44.05  & 40.92 & 39.34  & 38.11 & 37.48 &  39.98 & 41.95 \\
 & Reply Time & 6042.97 & 3036.12 & 1571.69 & 774.82 & 398.5 &  304.94 & 282.55 \\
 & Decrypt Time & 11.56 & 11.56 & 11.64 & 11.72 & 11.82 &  13.54 & 15.54 \\ 
 & \textbf{Total Time} & \textbf{6098.59} & \textbf{3088.59} & \textbf{1622.68} & \textbf{824.65} & \textbf{447.80} &  \textbf{358.47} & \textbf{339.66} \\ \midrule
 
\multirow{4}{*}{\textbf{Scenario 2b}} & Build Time &  24.55 & 20.4 & 19.2  & 18.3 & 18.1 &  19.75 & 21.95 \\
 & Reply Time & 3190.5 & 1618.3 & 808.05 & 410.35 & 215.8 &  159.25 & 168.65 \\
 & Decrypt Time & 11.05 & 11.25 & 11.15 & 11.05 & 12 &  13.15 & 15.35 \\ 
 & \textbf{Total Time} & \textbf{3227.35} & \textbf{1650.85} & \textbf{840.05} & \textbf{441.05} & \textbf{246.75} &  \textbf{193.3} & \textbf{207} \\ \midrule

\multirow{4}{*}{\textbf{Scenario 3}} & Build Time & 10.63 & 10.45 & 10.47 & 10.36 & 10.9 & 12 & 14.82 \\
 & Reply Time & 2044.14 & 1055.51 & 590.412 & 303.40 & 153.24 & 162.02 & 161.22 \\
 & Decrypt Time & 112.8 & 56.65 & 32.65 & 16.34 & 8.98 & 6  & 6.2 \\ 
 & \textbf{Total Time}  & \textbf{2167.58} & \textbf{1122.61} & \textbf{633.548} & \textbf{330.18} &  \textbf{173.13} & \textbf{190.06} & \textbf{192.28}\\ \midrule

%\multirow{4}{*}{\textbf{Scenario 3}} & Build Time & 18.84 & 18.50 & 18.57 & 18.50 & 19.72 &  20.9415 & 20.43 \\
% & Reply Time & 4153.99 & 2070.30 & 1147.99 & 603.36 & 305.167 &  313.918 & 321.37 \\
% & Decrypt Time & 68.69 & 86.82 & 52.38 & 41.23 & 13.1583 &  19.13 & 26.03 \\ 
% & \textbf{Total Time} & \textbf{4341.52} & \textbf{2175.62} & \textbf{1218.95} & \textbf{663.08} & \textbf{338.04} &  \textbf{353.98} & \textbf{367.83} \\ \midrule

%\multirow{4}{*}{\textbf{Scenario 3}} & Build Time & 18.27 & 18.20 & 18.71 & 18.30 & 19.16 &  21.27 & 22.16 \\
% & Reply Time & 4085.17 & 2040.27 & 1269.31 & 603.04 & 301.98 &  310.416 & 288.02 \\
% & Decrypt Time & 161.51 & 81.66 & 49.41 & 24.47 & 12.72&  19.1175  & 25.06 \\ 
% & \textbf{Total Time} & \textbf{4264.95} & \textbf{2140.13} & \textbf{1337.44} & \textbf{645.81} & \textbf{333.87} &  \textbf{350.8} & \textbf{335.44} \\ \bottomrule
\end{tabular}%
}
\end{table*}

\subsubsection{Scenario 2a: Single-Server (Batched)}
In this configuration, the complete multi-hop routing database is hosted by a single server, achieving computational efficiency through the batching mechanism of the BFV homomorphic encryption scheme. BFV supports Single-Instruction--Multiple-Data (SIMD) evaluation by representing plaintexts as vectors of $N$ slots, where $N$ equals the polynomial modulus degree ($N = 4096$).

Each database record consists of 27 hop identifiers placed into consecutive batching slots. Consequently, a single batched plaintext encodes $\lfloor \frac{4096}{27} \rfloor = 151$ independent database rows. The full database is therefore partitioned into $\lceil \frac{191577}{151} \rceil = 1269$ chunks. During query generation, the client constructs an encrypted one-hot selection vector where only the slots corresponding to the desired row are set to 1. The server performs a slot-wise ciphertext--plaintext multiplication for each chunk; the one-hot pattern suppresses non-target rows, ensuring only the target chunk produces a non-zero output. This strategy reduces the amortized per-row homomorphic cost by a factor of approximately 151 compared to the unbatched baseline.

\subsubsection{Scenario 2b: Single-Server (Optimized Packing)}
To further increase batching density, we implemented a compact encoding where two hop identifiers are packed into a single batching slot. Because hop values lie in the range $[0,630]$, we select a base $B = 1024$ and encode a pair $(h_0,h_1)$ as $x = h_0 + h_1 \cdot B$. As the BFV plaintext modulus is $t = 2^{20}$, the packed value satisfies $x < t$ and allows lossless decoding via $h_0 = x \bmod B$ and $h_1 = \lfloor x / B \rfloor$. This scheme reduces the storage per row to 14 slots, allowing a single plaintext to encode 292 rows and nearly halving the total number of required chunks.

\subsubsection{Scenario 3: Parallel Servers (Batched)}
We implemented a multi-server PIR scheme where the global database is vertically partitioned across 27 independent servers. Server $h$ stores only the $h$-th column (hop) for all rows. This construction combines the standard single-server model with the BFV SIMD batching described in Scenario 2a.

Each column is partitioned into blocks of $N$ rows. To query row $i$, the client calculates the block index $n^\star = \lfloor i/N \rfloor$ and the intra-block slot $\ell^\star = i \bmod N$. The client constructs a query vector of ciphertexts where only the ciphertext at index $n^\star$ encrypts a one-hot vector with a 1 in slot $\ell^\star$. This same query is sent to all 27 servers. Each server multiplies the query with its plaintext blocks in the NTT domain and sums the results to produce a single response ciphertext. Upon decryption, the client extracts the value at slot $\ell^\star$ to recover the specific hop for that column. This ensures standard PIR privacy, as the server processes all blocks identically and learns nothing about the queried index $i$.

\subsubsection{Scenario 4: Multi‑Query Vertical PIR (27‑Server Architecture)}
This scenario extends the 27‑server parallel PIR design to support high‑throughput evaluation of multiple independent queries. Instead of issuing a single query, the client constructs $m$ independent PIR queries, distributes them to all servers, and evaluates all $m\times$27 replies in parallel. The client decrypts each response to reconstruct $m$ complete 27‑hop paths. This approach preserves the original PIR trust and privacy model while significantly increasing throughput through parallelism across queries, servers, and ciphertext chunks.

\subsection{Performance Optimizations}
To achieve practical latencies, several key optimizations were implemented:
\begin{itemize}
    \item \textit{Symmetric Query Encryption:} The client uses its secret key to encrypt the query. This is significantly faster than public-key encryption and results in smaller ciphertexts, reducing query generation time and network overhead.
    \item \textit{NTT Pre-computation}: The server pre-processes its entire database into Number Theoretic Transform (NTT) form at startup. This enables the use of highly optimized multiplication algorithms during the homomorphic evaluation, addressing the primary computational bottleneck.
    \item \textit{Multi-threading:} Server-side computations, specifically the query-database multiplication loop, are parallelized across available CPU cores using OpenMP.
\end{itemize}

\begin{figure*}[htbp]
    \centering
    \includegraphics[width=\textwidth]{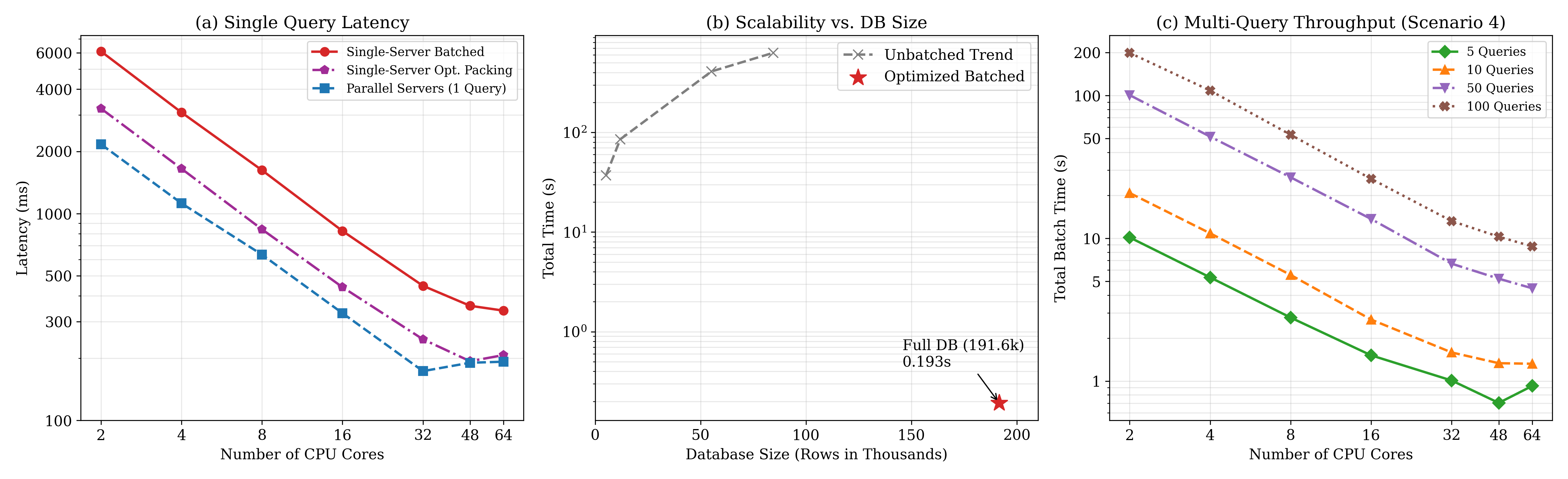}
   \caption{Comprehensive performance scalability analysis of the proposed PIR architectures.
    \textbf{(a) Single-Query Latency vs. Cores:} Comparison of optimized configurations. The parallel server architecture (Scenario 3, blue square) achieves the global minimum single-query latency of \SI{173.13}{\milli\second} at 32 cores, slightly outperforming the single-server optimized packing strategy (Scenario 2b, purple pentagon). Diminishing returns due to coordination overhead are observed beyond 32 cores.
    \textbf{(b) Scalability vs. Database Size:} The unbatched baseline (grey dashed line) exhibits untenable quadratic latency growth. In contrast, the optimized batched approach (red star) effectively decouples computational cost from database row count, processing the full \num{191577} row constellation in under \SI{200}{\milli\second}.
    \textbf{(c) Multi-Query Throughput (Scenario 4):} Total batch processing time for simultaneous query loads ($m=\{5, 10, 50, 100\}$) using the 27-server architecture. The system demonstrates strong scaling under heavy loads; at 64 cores, processing a batch of 100 queries takes \SI{8.77}{\second}, yielding a highly efficient amortized cost of approximately \SI{87.7}{\milli\second} per query.}
    \label{fig:pir_evaluation}
\end{figure*}

\subsection{Performance Evaluation}

The experimental evaluation was conducted on a high-performance compute node equipped with an AMD EPYC 7702P 64-Core Processor. We measured the total end-to-end latency, defined as the sum of client query generation (Build), server processing (Reply), and client decryption (Decrypt) times. The system was stressed across varying thread counts ranging from 2 to 64 cores to assess parallel scalability. The end-to-end processing times are summarized in Table \ref{tab:pir_detailed} and visualized in Figure \ref{fig:pir_evaluation}.

\subsubsection{Baseline Bottlenecks (Scenario 1)}
The baseline benchmarks illustrate the severe computational cost of classical, unbatched PIR. As shown in Figure \ref{fig:pir_evaluation}(b), the unbatched single-server configuration exhibits quadratic latency scaling with database size. Even with 64 cores, retrieving a short 5-hop path requires approximately \SI{37}{\second} (\SI{37024}{\milli\second}). Extending this to a partial path of 12 hops (\num{84421} rows) results in retrieval times exceeding \SI{629}{\second}, confirming that naive approaches are infeasible for real-time routing due to the lack of SIMD vectorization.

\subsubsection{Impact of Batching and Packing (Scenario 2a \& 2b)}
The introduction of SEAL-native batching (Scenario 2a) dramatically improves performance by packing the database into SIMD-compatible polynomials. This reduces the total latency for the full database to \SI{339.66}{\milli\second} at 64 cores. Further efficiency is gained in Scenario 2b, which utilizes an optimized packing strategy (2 hops per slot). This effectively halves the required homomorphic operations, lowering the minimum latency to \SI{193.3}{\milli\second} at 48 cores. As illustrated in Figure \ref{fig:pir_evaluation}(b), this optimized single-server approach effectively decouples computational cost from database size, processing the full \num{191577} row database in under \SI{200}{\milli\second}.

\subsubsection{Parallel Scalability} Scenario 3 leverages vertical partitioning to distribute a single query's workload across 27 parallel server instances. As shown in Figure \ref{fig:pir_evaluation}(a), this architecture achieves the global minimum single-query latency of \SI{173.13}{\milli\second} at 32 cores. However, both single-server and parallel architectures exhibit diminishing returns beyond this point. Scenario 3 saturates at 32 cores and slightly regresses to \SI{192.28}{\milli\second} at 64 cores. This behavior indicates that for a single query, once the per-core computational workload becomes sufficiently small, thread coordination and data aggregation overheads begin to dominate.

\subsubsection{Multi-Query Evaluation}  Scenario 4 evaluates the system's ability to process simultaneous batches of $m$ independent queries. Figure \ref{fig:pir_evaluation}(c) plots the total batch completion time against core count for varying workloads. The results demonstrate impressive scalability under load. While a single query faces coordination headwinds at 64 cores, a batch of 100 concurrent queries fully utilizes the available computational resources. At 64 cores, the total time to resolve 5 queries is \SI{0.92}{\second} (approx. \SI{184}{\milli\second}/query), whereas resolving 100 queries takes only \SI{8.77}{\second}. This yields a significantly improved amortized latency of approximately \SI{87.7}{\milli\second} per query. This result confirms that the proposed 27-server architecture is highly capable of supporting dense user traffic, becoming more efficient on a per-query basis as the simultaneous workload increases.

This sub-quarter-second delay for single query PIR is well within the acceptable threshold for session establishment or periodic route cache updates. Furthermore, the success of the consolidated architecture simplifies deployment by eliminating the complexity of managing a distributed 27-server cluster. 

It is important to note that the multi-query scenario 4 evaluated here represents a system-level parallelization of independent queries, rather than a specific Multi-Query PIR cryptographic construction where multiple requests are encoded into a single query ciphertext. While our current parallel approach confirms feasibility for medium-scale constellations, future work addressing global scalability (10,000+ satellites) will focus on implementing true cryptographic query batching. By encoding multiple route requests into a single ciphertext, we aim to further reduce communication overhead and amortize server-side processing costs. Additionally, we will explore advanced multi-server PIR schemes \cite{davidson2022frodopir, spiral, mulpir, onionpir} to ensure performance scales linearly with the growing database size.

\section{Performance Analysis: Enhanced Reliability with FEC}
\label{sec:evalfecperf}

To combat packet loss from topological volatility, we employ a multi-path transmission scheme based on Reed-Solomon FEC.

\begin{figure}[ht!]
    \centering
    \includegraphics[width=\columnwidth]{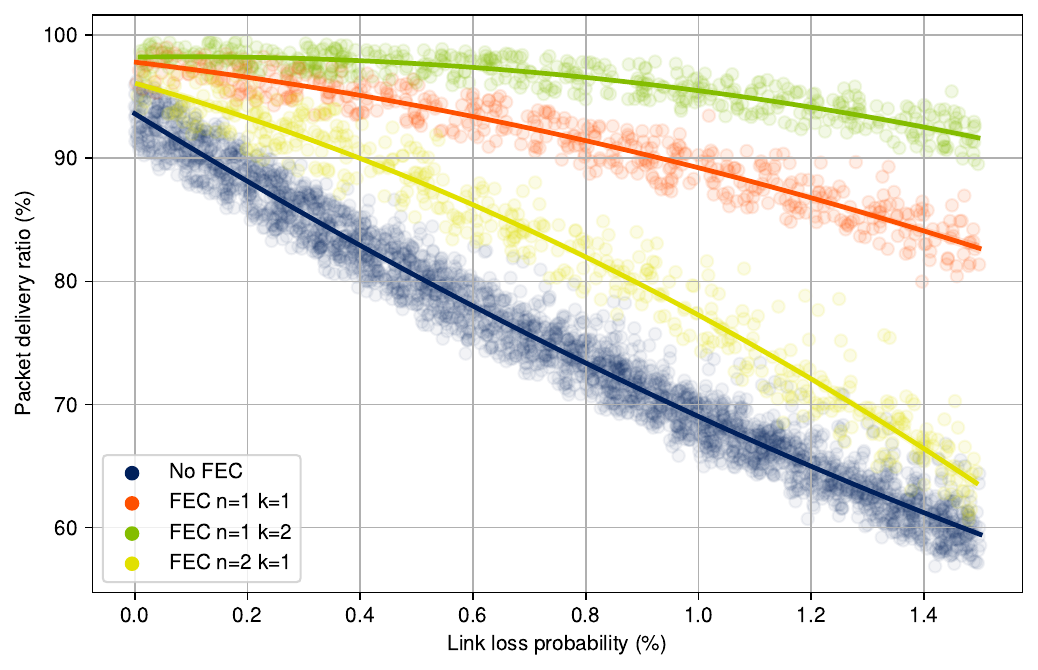}
    \caption{Packet delivery ratio as a function of link loss probability for different FEC configurations. The (n=2, k=1) scheme provides near-perfect delivery.}
    \label{fig:fec_reliability}
\end{figure}

\textit{Reliability vs. Overhead:} The most critical result is the demonstrated trade-off between reliability and bandwidth. As shown in Figure \ref{fig:fec_reliability}, link failures can significantly degrade the packet delivery ratio. Without FEC, a mere 1.5\% link loss probability results in a message loss rate of nearly 40\%. However, by encoding the message and sending it over multiple paths, reliability is dramatically improved. Sending one parity chunk alongside the original data ($n=2, k=1$) virtually eliminates message loss even with significant link instability.

\begin{figure}[ht!]
    \centering
    \includegraphics[width=\columnwidth]{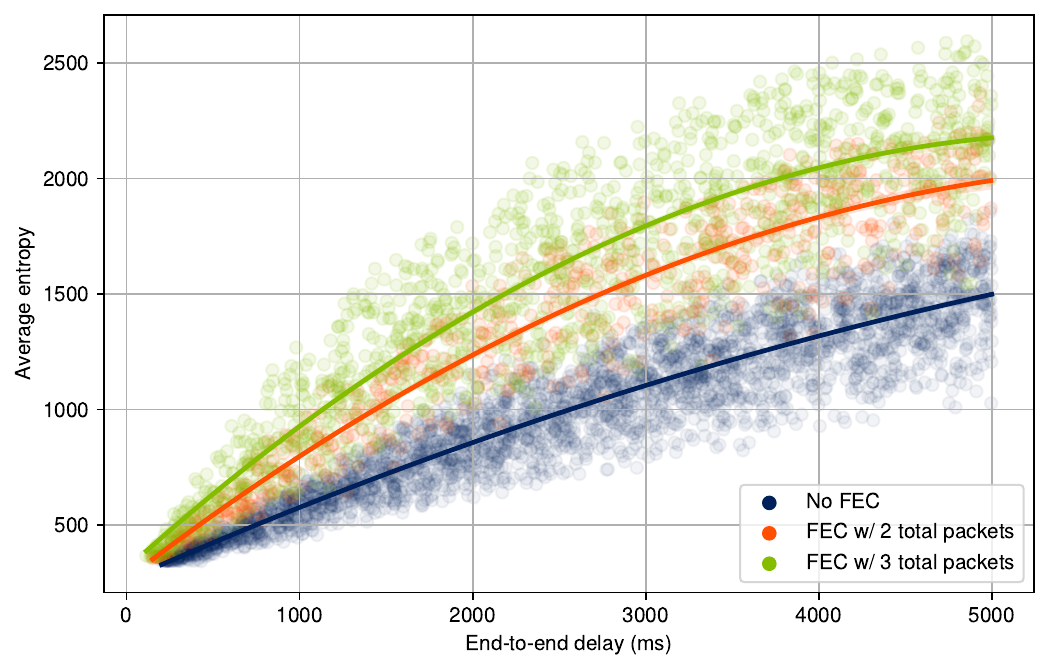}
    \caption{Comparison of average entropy versus end-to-end delay, showing that multi-path transmission with FEC enhances anonymity.}
    \label{fig:fec_anonymity}
\end{figure}

\subsection{Anonymity Benefits of FEC:} Beyond reliability, the multi-path nature of FEC provides a secondary benefit of enhanced anonymity. Splitting a single logical message across multiple, uncorrelated paths adds noise to the network, making it more difficult for an adversary to perform traffic analysis. As shown in Figure \ref{fig:fec_anonymity}, for any given end-to-end delay, using FEC with 2 or 3 total packets consistently achieves a higher average entropy than sending a single packet without FEC. This demonstrates that the mechanism for reliability also hardens the system against traffic correlation attacks.

\subsection{Computational Feasibility:} 
\begin{figure}[h!]
	\centering
	\includegraphics[width=\columnwidth]{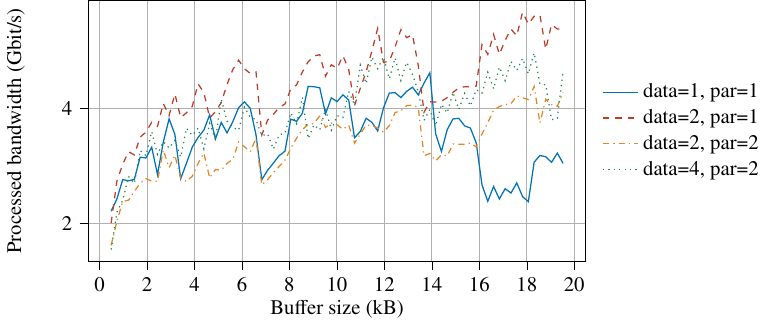}
	\caption{Performance of FEC encoding with different configurations}
	\label{fig:eval_fec_writer}
\end{figure}

\begin{figure}[h!]
	\centering
	\includegraphics[width=\columnwidth]{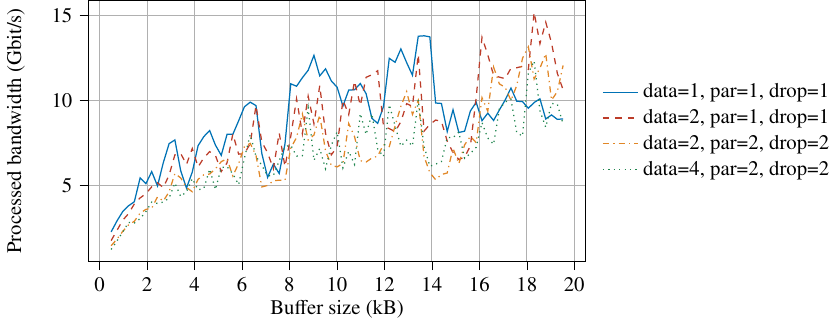}
	\caption{Performance of FEC decoding with different configurations}
	\label{fig:eval_fec_reader}
\end{figure}

Finally, to validate the computational feasibility of our reliability mechanism, we benchmarked the FEC encoding and decoding functions on our testbed hardware, meaning the performance of the generation of parity data for the encoding, and the performance of the decoding when packets are lost. As shown in Figure  \ref{fig:eval_fec_writer} and \ref{fig:eval_fec_reader}, the Reed-Solomon encoder achieves an average throughput of approximately 4 Gbit/s, while the decoder, which benefits from being able to reconstruct data in parallel, achieves approximately 8 Gbit/s. These high-throughput results demonstrate that the computational cost of FEC is not a bottleneck and is entirely practical for provider nodes and ground stations to manage. 

\subsection{Countering Traffic Centralization with Adaptive Delays}
To mitigate the traffic centralization at polar nodes, we explored adaptive delay strategies where mix nodes introduce delays based on their topological position, forcing traffic to mix more effectively at these choke points. We evaluated two such strategies:
\begin{enumerate}
    \item \textit{Centrality-based delay:} Nodes with higher network centrality are configured to introduce larger delays.
    The following formula is used for computing the delay:
    \begin{equation}
    	\textrm{node delay} = \frac{\textrm{delay parameter}}{1 + \textrm{node betweenness centrality}}
    \end{equation}
    
    \item \textit{Latitude-based delay:} Nodes at higher latitudes (closer to the poles) introduce smaller delays.
    The following formula is used for computing the delay:
    \begin{equation}
    	\textrm{node delay} = \textrm{delay parameter} \cdot \left(1 - \frac{|\textrm{node latitude}|}{90}\right)
    \end{equation}
\end{enumerate}

\begin{figure}[ht!]
    \centering
    \includegraphics[width=\columnwidth]{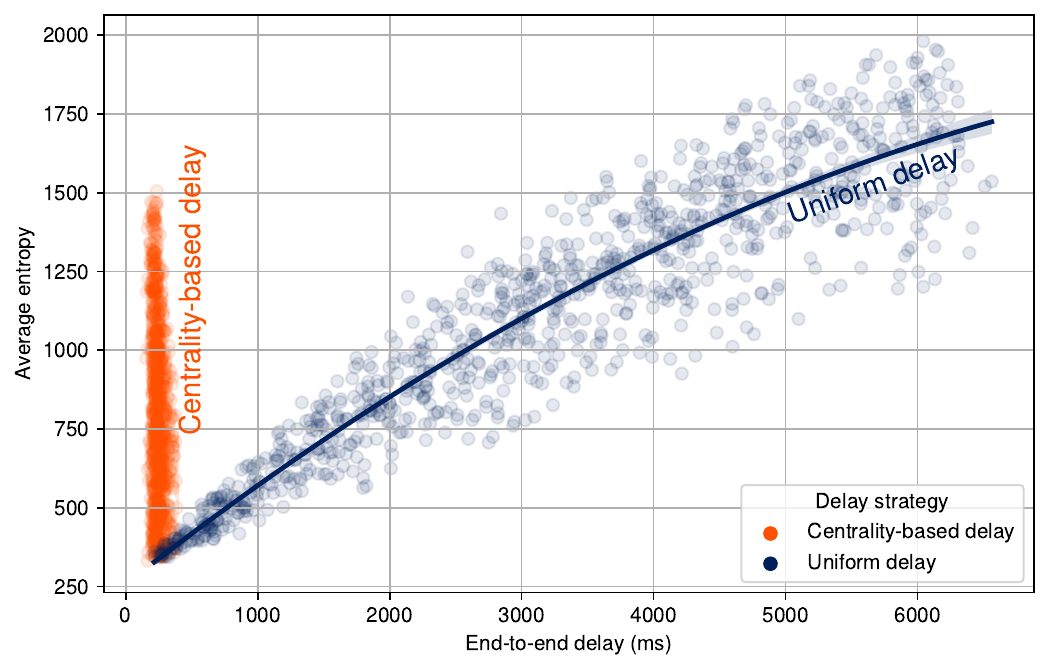}
    \caption{Comparison of entropy vs. delay for Uniform and Centrality-based delay strategies.}
    \label{fig:delay_strategy}
\end{figure}

\begin{figure}[ht!]
    \centering
    \includegraphics[width=\columnwidth]{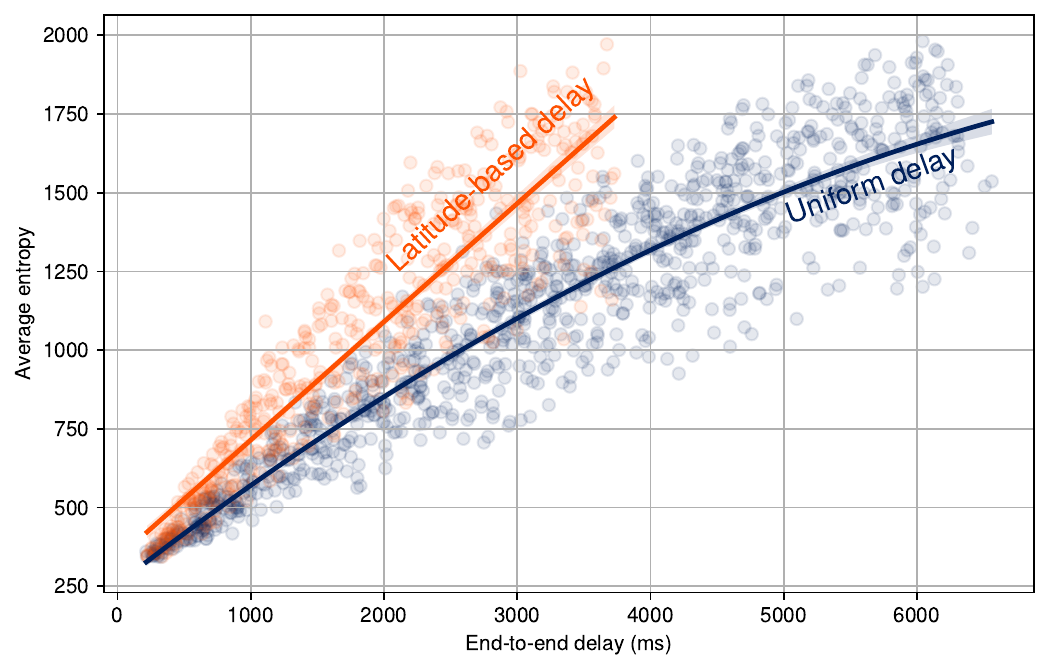}
    \caption{Comparison of average entropy versus end-to-end delay for Latitude-based and Uniform delay strategies.}
    \label{fig:latitude_delay}
\end{figure}

As shown in Figure \ref{fig:delay_strategy}, the centrality-based strategy is highly effective, achieving a high level of entropy within a narrow and low range of end-to-end delays. Similarly, Figure \ref{fig:latitude_delay} shows that the latitude-based delay strategy also provides a superior entropy-latency trade-off compared to a uniform delay. Both strategies demonstrate that by strategically concentrating the mixing effort at the network's natural choke points, we can achieve strong anonymity guarantees more efficiently and counter the topological bias inherent in the constellation.

\section{Analysis of Core Anonymity Parameters}
Finally, we analyzed fundamental system parameters that influence the core trade-off between anonymity and performance
\subsection{Impact of Path Selection Strategy}
In a source-routed anonymity system, the client's choice of path through the network significantly impacts both performance and privacy. We evaluate two distinct path selection strategies: selecting the \textit{shortest paths} between sender and receiver in terms of hop count, and selecting \textit{random paths} of a fixed length.

\begin{figure}[ht!]
    \centering
    \includegraphics[width=\columnwidth]{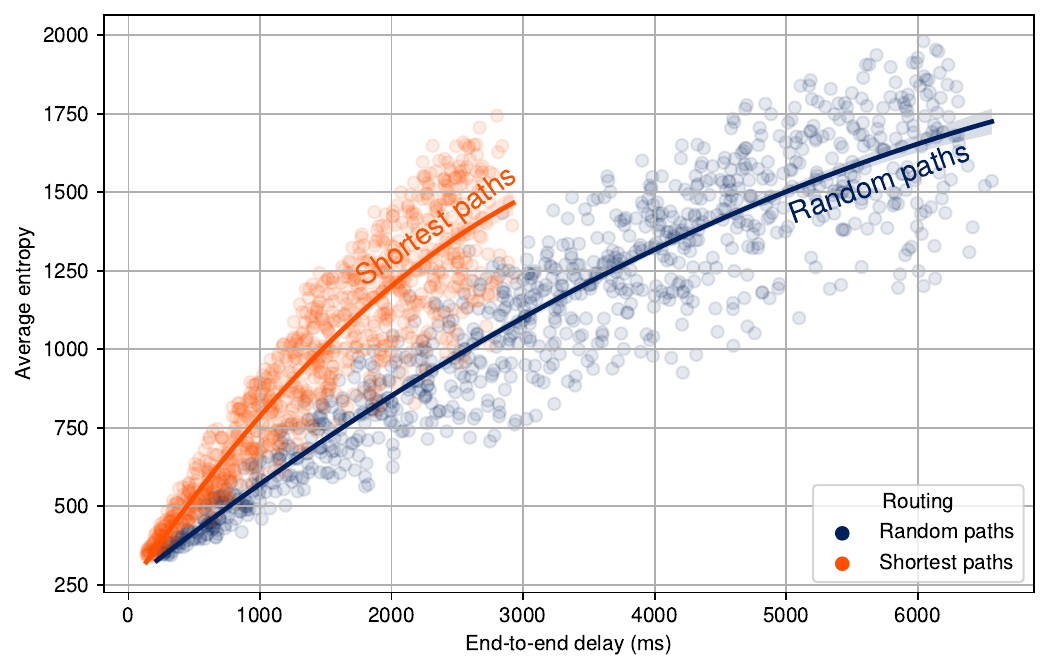}
    \caption{Comparison of average entropy versus end-to-end delay for Shortest Path and Random Path selection strategies.}
    \label{fig:routing_strategy}
\end{figure}

Figure \ref{fig:routing_strategy} illustrates a clear trade-off. Using the shortest paths naturally results in lower end-to-end delays but also yields lower entropy. Conversely, selecting paths randomly leads to longer delays but provides a significantly larger anonymity set for any given delay value. This is because random paths are less predictable and traverse a more diverse set of mix nodes, increasing the effectiveness of the mixing process. This demonstrates that sacrificing some latency by avoiding topologically obvious paths can lead to substantial gains in anonymity.

\subsection{Impact of Mix Buffer Size}
Exponential mixing involves holding packets in a buffer before forwarding them. The minimum size of this buffer can influence the anonymity provided. We evaluated the system with minimum buffer sizes ranging from 1 to 4.

\begin{figure}[ht!]
    \centering
    \includegraphics[width=\columnwidth]{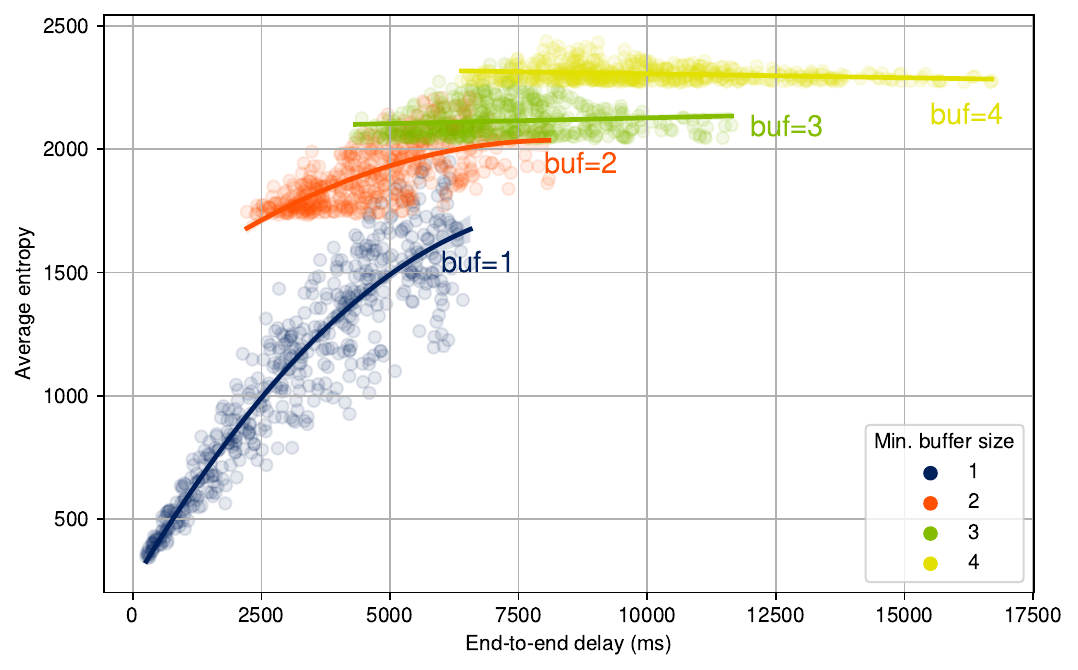}
    \caption{Impact of the minimum mix buffer size on the entropy-delay trade-off.}
    \label{fig:buffer_size}
\end{figure}

Figure \ref{fig:buffer_size} shows that increasing the minimum buffer size consistently improves the achievable anonymity for a given delay. A larger buffer allows a mix node to create a larger anonymity set for each output packet. However, the gains exhibit diminishing returns. The improvement from buf=1 to buf=2 is significant, but the improvement from buf=3 to buf=4 is much smaller, while the overall latency increases. This indicates that there is an optimal buffer size beyond which the additional latency costs may outweigh the marginal gains in privacy.
\begin{figure*}[htbp]
\centering
\includegraphics[width=0.8\textwidth]{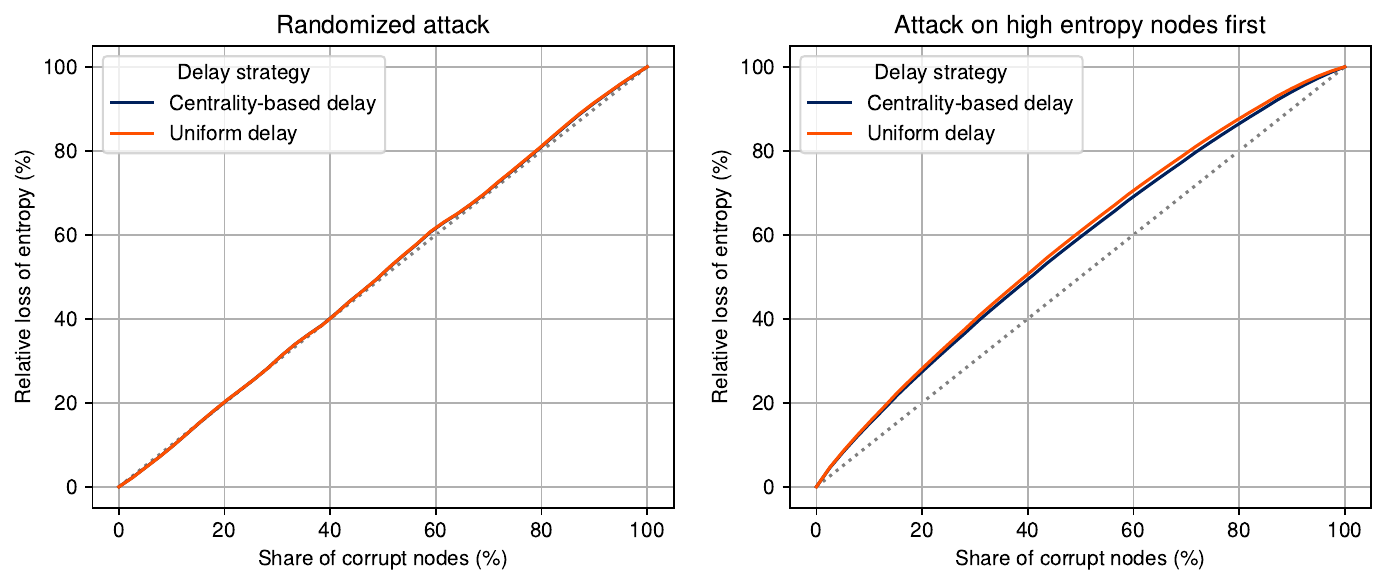}
\caption{Relative loss of entropy as a function of the percentage of compromised nodes under both a randomized attack and a targeted attack on high-entropy nodes.}
\label{fig:node_compromise}
\end{figure*}

\subsection{Resilience to Node Compromise} 
A crucial measure of a system's robustness is how its anonymity degrades as an adversary gains control over network nodes. Figure \ref{fig:node_compromise} evaluates the system's resilience against two distinct adversarial strategies: a randomized attack, where the adversary compromises nodes uniformly at random, and a targeted attack, where the adversary strategically compromises the most important nodes (those with the highest entropy) first.

The results show a stark difference between the two strategies. Under a randomized attack, the system exhibits graceful degradation. The relative loss of entropy is nearly linear with the share of corrupt nodes, meaning that an adversary must compromise a significant fraction of the network to cause a proportional loss of anonymity. This aligns with the theoretical model, which predicts a gradual decline in privacy as random nodes are compromised.

In contrast, the targeted attack is significantly more effective. By focusing on the high-entropy nodes that contribute most to the system's anonymity, the adversary can cause a much more rapid loss of privacy. For example, compromising just 20\% of the most critical nodes results in a nearly 30\% loss of entropy. This demonstrates that while the system is robust against random failures or an unsophisticated adversary, a strategic adversary with knowledge of the network's traffic patterns can degrade its privacy guarantees much more efficiently.

%\subsubsection{Incremental Entropy at a Mix Node}
%The core trade-off in a mix-network is between the anonymity it provides and the latency it introduces. Anonymity, measured as the entropy of the mix pool, is generated by collecting and mixing traffic from diverse sources. This process inherently takes time.

%The anonymity of the system is not static; it evolves with the network conditions. Figure \ref{fig:entropy_evolution} shows the change in pool entropy for a mix-node over time for different numbers of provider nodes (pn) and traffic rates (d). The results clearly indicate that both higher user participation (more providers) and higher traffic rates lead to a greater and more stable level of entropy. With 1000 providers and a delay of of 500ms, the system achieves a high and consistent entropy level, demonstrating that the system's anonymity properties scale well with the size and activity of its user base.

%\begin{figure}[ht!]
%    \centering
%    \includegraphics[width=\columnwidth]{plots/Change in the entropy of the pool over time.png}
%    \caption{Change in pool entropy over time for different numbers of providers (pn) and average delay (d) of message per node.}
%    \label{fig:entropy_evolution}
%\end{figure}

\subsection{Impact of Packet Mixing Strategy}
The core of a mix-network is its strategy for delaying and reordering packets. We compare two common strategies: \textit{Stop-and-go mixing}, where a mix node holds packets until a certain number have been collected or a timeout is reached, and \textit{Exponential mixing} (also known as Poisson mixing), where each packet is delayed by a random duration drawn from an exponential distribution.

\begin{figure}[ht!]
    \centering
    \includegraphics[width=\columnwidth]{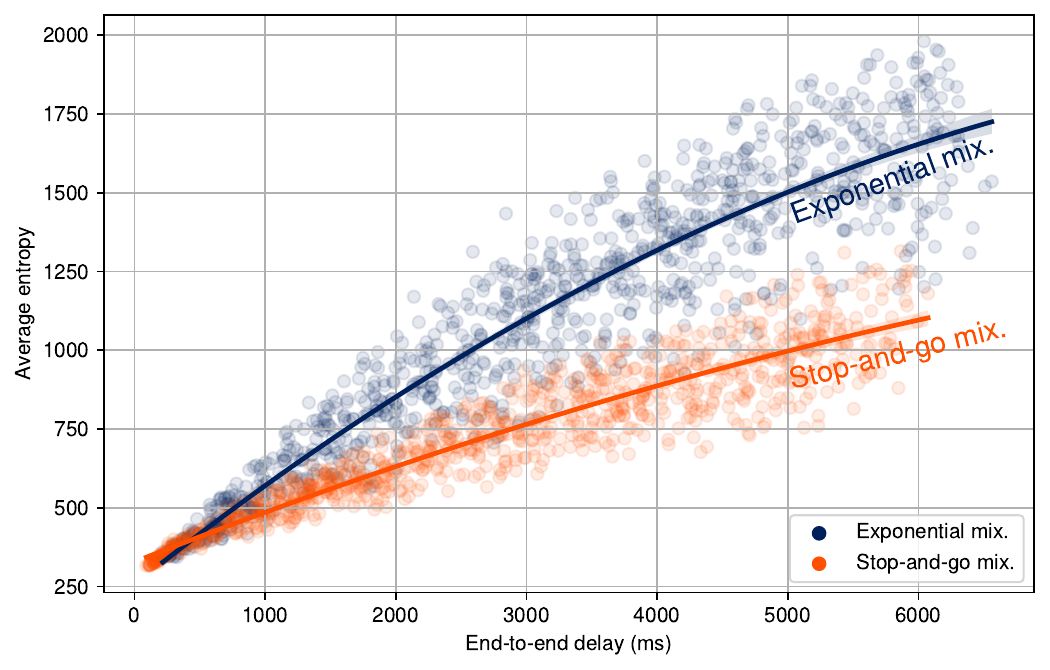}
    \caption{Comparison of average entropy versus end-to-end delay for Exponential and Stop-and-go mixing strategies.}
    \label{fig:mixing_strategy}
\end{figure}

As shown in Figure \ref{fig:mixing_strategy}, for any given end-to-end delay, the exponential mixing strategy consistently achieves a higher average entropy than the stop-and-go strategy. The random, memoryless nature of the exponential delay is more effective at confusing a timing-based adversary. This result confirms that exponential mixing provides a more efficient trade-off, delivering better anonymity for the same latency cost.

\section{Conclusion}

This work details a comprehensive architecture designed to provide robust and private communications within highly dynamic network environments, such as LEO satellite constellations. By extending the Loopix anonymity system, our design directly confronts the dual \textit{challenges of path unreliability} due to topological volatility and \textit{metadata leakage during route discovery}.

The evaluation demonstrates the effectiveness of the two primary contributions. First, the integration of a \textit{multi-path transmission scheme} using erasure codes (ie. FEC) not only makes the system resilient to packet loss—virtually eliminating message loss even with significant link instability — but also enhances anonymity by splitting traffic across uncorrelated paths.
Second, the application of a \textit{privacy-preserving route discovery protocol} using homomorphic encryption (using PIR) is shown to be practically feasible. The optimized implementation achieves a total query latency of under two seconds against a large, real-world database, successfully closing a critical metadata leakage channel without imposing prohibitive delays for session initializations or periodic route updates.

Furthermore, the analysis confirmed that adaptive delay strategies, particularly those based on network centrality, can efficiently counteract the inherent topological biases of LEO constellations, yielding stronger anonymity for lower latency compared to uniform approaches. The system exhibits graceful degradation against random node compromise, although it remains vulnerable to sophisticated, targeted attacks on high-entropy nodes.

The LEO satellite network proved to be a prime use case for this architecture. The inherent challenges of the environment — namely the constant topological change and the non-uniform traffic distribution concentrating on polar nodes — were effectively mitigated by the proposed countermeasures. A key advantage of this use case is the ability to leverage the satellite network's public, high-volume data stream as a massive source of cover traffic. This allows the system to embed private communications within the public data flow, achieving a high degree of anonymity and unobservability in a cost-effective manner by reducing the need to generate dedicated cover traffic. The robust security model is particularly relevant for governmental or military communications that may traverse various jurisdictions and be subject to widespread surveillance.

Building on these findings, several avenues for future research emerge: the current system relies on static topology snapshots. Future iterations could integrate predictive algorithms that leverage deterministic satellite orbital mechanics. By selecting paths with a higher probability of long-term stability, such schemes could minimize route failures and significantly reduce the bandwidth overhead currently required by Forward Error Correction (FEC).

Although current query latencies are practical, further optimization is desirable. Future work should benchmark alternative protocols: schemes like OnionPIR \cite{onionpir} and SpiralPIR \cite{spiral} offer logarithmic communication costs, benefiting bandwidth-constrained links. Conversely, preprocessing-based protocols like SimplePIR \cite{simple} and FrodoPIR \cite{frodo} could significantly boost computational throughput. Additionally, exploring hardware acceleration via GPUs or FPGAs, alongside Trusted Execution Environments (TEEs), presents a promising path for reducing server-side processing times.
 
Our evaluation assumed a passive global adversary with static node compromises. Future analysis must extend to sophisticated, active adversaries capable of adaptive surveillance or disruption attacks, such as targeted link flooding, to rigorously test the system's resilience in adversarial environments.

%A key industrial application of our work is the development of a shared, EU-based LEO constellation. This dual-use approach is highly cost-efficient due to its large number of commercial users, and our protocol provides the high degree of anonymity essential for its state and military users.

\bibliographystyle{IEEEtran}
\bibliography{sample-base}

\end{document}